\documentclass{PoS}

\title{The contribution of hard processes to elliptic flow}

\usepackage{amsmath}

\ShortTitle{The contribution of hard processes to elliptic flow}

\author{\speaker{Boris Tom\'a\v{s}ik}\thanks{Work supported in parts by 
VEGA 1/4012/07 (Slovakia), MSM~6840770039 and  LC~07048 (Czech Republic).
The participation to the workshop has been made possible through Hungaro-Slovak 
exchange grant SK-MAD 02906.}\\
        Univerzita Mateja Bela, Bansk\'a Bystrica, Slovakia\\
				Faculty of Nuclear Science and Physics Engineering, Czech Technical University in Prague, 
				Prague, Czech Republic\\
        E-mail: \email{tomasik@fpv.umb.sk}}

\abstract{%
I study a possible effect of momentum  deposition from many hard 
partons traversing the hot and dense region produced early in nuclear 
collisions at the LHC. The expected number of such hard partons is large.
It is argued that the induced diffusion wakes which carry the momentum
deposited to the medium may interact. Due to azimuthally asymmetric geometry 
in non-central collisions, this may lead to small preference of the collective flow 
in the direction of the impact parameter. As a result, a small contribution 
to the azimuthal asymmetry of hadronic spectra is obtained. It may be 
important to take even such a small contribution into account if 
quantitative conclusions about early thermalisation and low viscosity 
are to be made, based on the measurements of the elliptic flow parameter $v_2$.
}

\FullConference{High-pT Physics at LHC - Tokaj'08\\
		 March 16 - 19 2008\\
		 Tokaj, Hungary}

\begin{document}

\section{Introduction: elliptic flow}

Elliptic flow is the name for an azimuthal asymmetry of hadron production in non-central 
relativistic heavy ion collisions. The name is suggested by the most common interpretation: the bulk
matter excited to high energy densities expands in the directions transverse to the beam due to 
pressure gradients. In non-central collisions, the overlap region of the 
two nuclei is not symmetric in azimuthal angle, but rather almond-shaped with the shorter
size in the direction of the impact parameter. As the pressure gradients are larger where 
the size of the fireball is shorter, stronger transverse expansion will be generated in the
direction of the impact parameter. Through the Doppler effect, more particles and flatter 
spectra are then emitted in the direction of stronger expansion and this leads to the 
observed azimuthal asymmetry of the spectra. Hence the name elliptic flow. 

Elliptic flow is measured in terms of the second order Fourier coefficient of the azimuthal distribution 
of produced hadrons, $v_2$. This is introduced as
\begin{equation}
E \frac{d^3N}{dp^3} = \frac{d^2N}{p_t\, dp_t\, dy}\, \frac{1}{2\pi}\, \left ( 1 + 
2v_2(y,p_t)\cos\left ( 2(\phi - \phi_R)\right ) 
+ \dots \right )
\end{equation}
where $\phi_R$ is the azimuthal angle of the \emph{reaction plane}, defined by the beam axis and the impact parameter.
Note that other  than even cosine terms vanish  in symmetric collisions at midrapidity for symmetry 
reasons. 

Observed at RHIC, the elliptic flow is rather large. In hydrodynamic simulations, it turns 
out that such a  large flow asymmetry can be achieved only if very fast thermalisation 
is assumed \cite{Heinz:2001xi,Pratt:2008sz,Broniowski:2008vp} and the shear viscosity is extremely low \cite{Teaney:2003kp}.
These are important conclusions from RHIC, though their full theoretical understanding is 
lacking yet. However, in order to ensure the reliability of these conclusions one should 
understand the interpretation of the elliptic flow very well. Particularly, all effects which cause elliptic flow 
in addition to the expansion due to pressure gradients should be considered. In this contribution I 
examine one such possible effect. 

%%%%%%%%%%%%%%%%%%%%%%%%%%%%%%%%%%%%%%%%%%%%%%%%%%%%%%%%%%%%%%%%%%
\section{Introduction: quenching of hard partons}

In nuclear collisions at highest energy, such as those at the LHC, but to some extent 
also at RHIC, hard scatterings
between incident partons happen frequently. They normally lead to production of jets or minijets. However, rather 
few of these jets are indeed observed. The majority of the leading hard partons loose energy totally
in the surrounding strongly interacting medium. On the other hand, this also means that the energy and momentum 
are transferred to the bulk matter and it is appropriate to ask about how the matter 
responds. The response of the bulk is currently widely discussed in the literature \cite{Stoecker:2004qu,Satarov:2005mv,CasalderreySolana:2004qm,Ruppert:2005uz,Betz:2008js}. 
It is important to note, however, that any such response is primarily correlated with the direction 
of the inducing jet and not with the direction of the reaction plane. A possible correlation with the 
reaction plane may be due to different energy loss in various direction of the initial jet. 

At the LHC,  we expect many such (mini)jets to be present in a single collision. What is the effect 
on the bulk medium of momentum deposition of all of them? Since the directions of their original velocities 
are distributed isotropically, the simplest expectation would be that after summing up all momentum depositions 
we end up with transversally isotropic flow. The situation is less clear, however, in non-central collisions. 
The asymmetry of sizes between the direction in the reaction plane  
and that out of the reaction plane may be reflected in an asymmetry of the collective effect 
resulting from momentum deposition from hard partons. Such an effect is examined here.

%%%%%%%%%%%%%%%%%%%%%%%%%%%%%%%%%%%%%%%%%%%%%%%%%%%%%%%%%%%%%%%%%%
\section{The effect of many jets}

\begin{figure}[t]
\begin{center}
\includegraphics[width=0.6\textwidth]{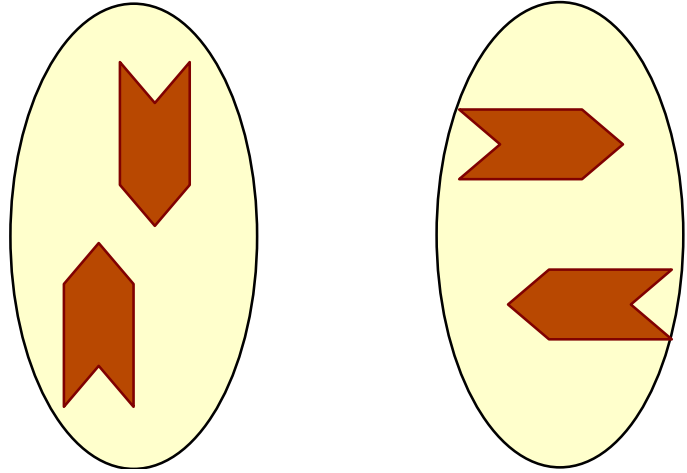}
\end{center}
\caption{%
Illustration of the probability that two streams could meet. Left: two streams 
flowing in the out-of-plane direction have a better chance to meet. Right: as the fireball 
is elongated out of the reaction plane, two streams which flow in the in-plane direction 
have more space to pass each other.
}
\label{f:cart}
\end{figure}

The main idea here will be that the jets induce some kind of streams when they deposit momentum 
into the bulk. In the literature, they are known as diffusion wakes. It is important that even if the leading 
partons are fully stopped, the streams seem to continue and carry momentum \cite{betz}.
Originally, they flow in all transverse directions isotropically. Let us imagine, however, that 
two such streams come together from exactly opposite directions. Their momenta would \emph{cancel} and 
energy would be deposited into the place of the merger. Thus the flow-generating effect would be smaller than 
just from a simple addition of two streams.
In a very handwaving and cartoon-like way one could argue that in non-central collisions there is better 
chance of this to happen if the jets fly in the direction perpendicular to the reaction plane. The situation 
is sketched in Figure~\ref{f:cart}. 
%%%%%%%%%%%%%%%%%%%%%%%%%%%%%%%%%%%%%%%%%%%%%%%%%%%%%%%%%%%%%%%%%%%%%%
%%%%%%%%%%%%%%%%%%%%%%%%%%%%%%%%%%%%%%%%%%%%%%%%%%%%%%%%%%%%%%%%%%%%%
The fireball is elongated out of the reaction plane. Thus two jet-generated streams of bulk matter may 
have enough space to pass each other and not cancel, if they fly in the reaction plane. Then they 
both lead to observable asymmetry in the azimuthal
hadron distribution. Streams which have their direction perpendicular to the reaction plane have 
less space available for passing by and the probability  that they will merge and their momenta
will cancel is higher. 

This reasoning would suggest, that the effect of having many hard partons inducing flow in the fireball 
and the individual flows merging would lead to a positive net contribution to the elliptic flow parameter
$v_2$. The proper way to test this conjecture would be a hydrodynamic simulation with the jets 
feeding the flow, i.e.\ technically source terms for in the hydrodynamic equations would be introduced. 
At the moment, such 
a simulation is technically too complicated with very large number of jets. I therefore choose a simpler way 
and construct a toy model to represent the situation under study.

%%%%%%%%%%%%%%%%%%%%%%%%%%%%%%%%%%%%%%%%%%%%%%%%%%%%%%%%%%%%%%%%%%
\section{The toy model}

The streams within the fluid are represented by blobs  of matter. They all fly 
with velocities of 0.999$c$ in various directions. Below it is described how the directions 
and initial positions are chosen. When two blobs meet, 
they merge into one which has larger mass and the 
momentum  such that energy and momentum are conserved. In this way the merger of 
two streams (diffusion wakes) is represented. In the end, when there are no more mergers,
the blobs  evaporate pions according to a thermal distribution with a temperature
of 170~MeV until all their energy is used up. 

Here we make a calculation for the LHC. 

The blobs carry momentum according to the distribution of hard partons in 
transverse energy and pseudorapidity, which is parametrised as 
\begin{eqnarray}
\label{Etsig}
\frac{d\sigma_{NN}}{dE_T} & = & 8.3385 \cdot 10^8  \left ( \frac{E_T}{1\,\mbox{GeV}} \right )^{-4.29717} 
\mu\mbox{b/GeV} \\
\label{etasig}
\frac{d\sigma_{NN}}{d\eta} & \propto &  1 - 0.067017 \eta^2\, .
\end{eqnarray}
The normalisation here is for single nucleon-nucleon collision. 
The differential cross section in $E_T$ is integrated over the pseudorapidity 
interval [--2.5,2.5]. It is assumed that the distributions of jets in $E_T$ and 
$\eta$ factorize in this interval. These parametrisations have been obtained from 
fits to MC results published in \cite{Accardi:2004gp}. There, they were shown for $E_T$ above 
20~GeV. Here they will be extrapolated to lower $E_T$'s. As a result, the multiplicity
of jets will be slightly overestimated. A more realistic parametrisation is being worked out. 

The azimuthal angle is distributed \emph{isotropically}. 
The mass of the blobs  is determined 
from the fixed velocity of 0.999$c$ and the generated momentum. 

The total average number of blobs in one collision is given by the number of (mini)jets
produced. This can be calculated with the help of parametrisations \eqref{Etsig}
and \eqref{etasig}. 
We first define the cross-section for a production of jets with transverse energy 
bigger than $E_m$
\begin{equation}
\sigma(E_m) = \int_{E_m}^{\infty} \frac{d\sigma_{NN}}{dE_T} \, dE_T\, ,
\end{equation}
and then determine the number of jets produced in the non-central nuclear collision 
of nuclei with mass number $A$ as
\begin{equation}
N_j(E_m,b) = \frac{A^2 \, T_{AA}(b)\, \sigma(E_m) \, K}{1 - \left (  1 - T_{AA}(b) \sigma(E_m) K  \right )^{A^2}}\, .
\end{equation}
In the last equation we have introduced the impact parameter $b$ and the nuclear overlap function
\begin{equation}
T_{AA}(b)  =  \int_{\rm overlap}  T_A(\vec r)\, T_A(\vec r - \vec b)\, d^2\vec r \, ,
\label{nuover}
\end{equation}
which is defined with the help of the nuclear thickness function
\begin{equation}
T_A(\vec r)  =  2\, \rho_0\, \sqrt{R_A^2 - r^2} \, .
\label{tadef}
\end{equation}
Integration in eq.~\eqref{nuover} goes over the whole overlap region of the two nuclei. In eq.~\eqref{tadef},
$\rho_0$ is the nuclear density; it is assumed here that the density is uniform. Note also the factor
$K = 10^{-4}$~fm$^2$/$\mu$b which converts the microbarns from the cross section parametrisation into fermis.

The original transverse positions of the blobs are generated from the distribution 
\begin{equation}
\rho(\vec r) = T_A(\vec r) T_A(\vec r - \vec b)\, ,
\end{equation}
which 
corresponds to scaling with the number of binary collisions. In the longitudinal direction and time, the space-time
point of their appearance is determined as
\begin{eqnarray}
x_0 & = & \tau_0 u_0 \\
x_3 & = & \tau_0 u_3\, ,
\end{eqnarray}
where $u_\mu$ is the four-velocity of the blob and $\tau_0$ is the formation time of the stream, 
which is chosen here 0.6~fm/$c$ in accord with the fast thermalisation conjecture \cite{Heinz:2001xi}.

The size of the blobs can be varied in order to simulate various transverse sizes of the streams. 
It is decisive for the collisions because it actually gives the size of the cross section for 
blob-blob mergers.

%%%%%%%%%%%%%%%%%%%%%%%%%%%%%%%%%%%%%%%%%%%%%%%%%%%%%%%%%%%%%%%%%%

\section{Results}

In the simulation from which results are presented the choice of parameters was
$E_m = 4$~GeV and the radius of the blobs 2.5~fm. Results were simulated in five 
centrality classes, which are summarised in Table~\ref{t:centr}.
%%%%%%%%%%%%%%%%%%%%%%%%%%%%%%%%%%%%%%%%%%%%%%%%
\begin{table}
\begin{center}
\begin{tabular}{|ccc|}
\hline\hline
centrality & $b$ [fm] & $N_j$ \\
\hline
0--10\% & 2.8 & 14174 \\
10--30\% & 5.9 & 7820 \\
30--50\% & 8.4 & 3330 \\
50--75\% & 10.5 & 933 \\
75--100\% & 12.4 & 56 \\
\hline\hline
\end{tabular}
\end{center}
\caption{The centrality classes which were used in the simulation. For each class
the impact parameter $b$ and the average multiplicity of minijets with $E_T > 4$~GeV 
in the simulated interval $N_j$ is 
presented.
\label{t:centr}}
\end{table}
%%%%%%%%%%%%%%%%%%%%%%%%%%%%%%%%%%%%%%%%%%%%%%%%
As mentioned and argued  in the previous section, the jet multiplicities are most 
probably overestimated and  better parametrisation of their production cross section 
is being elaborated. For the time being, let us proceed with the available parametrisation. 

\begin{figure}[t]
\begin{center}
\includegraphics[width=0.48\textwidth]{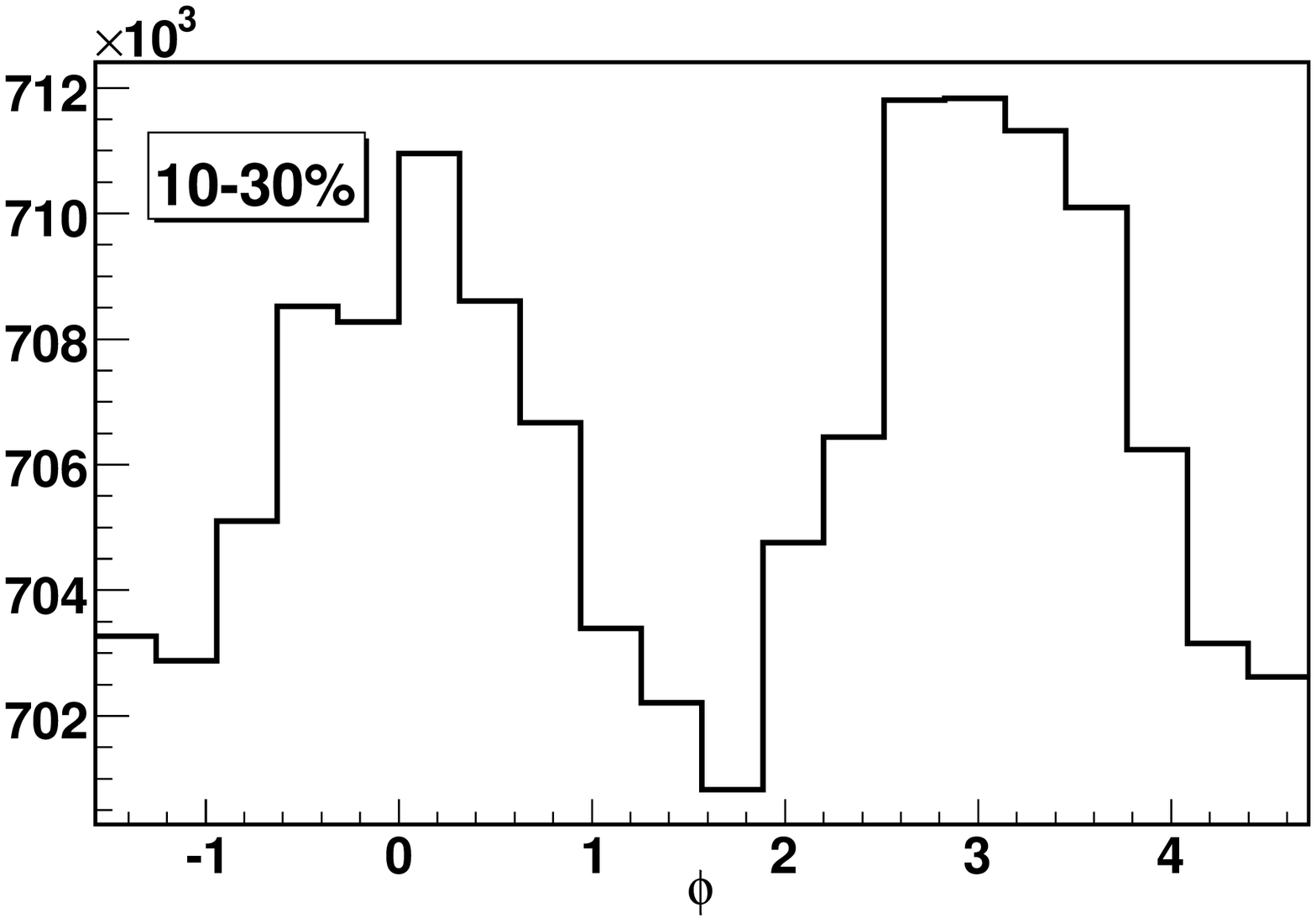}
\ 
\includegraphics[width=0.48\textwidth]{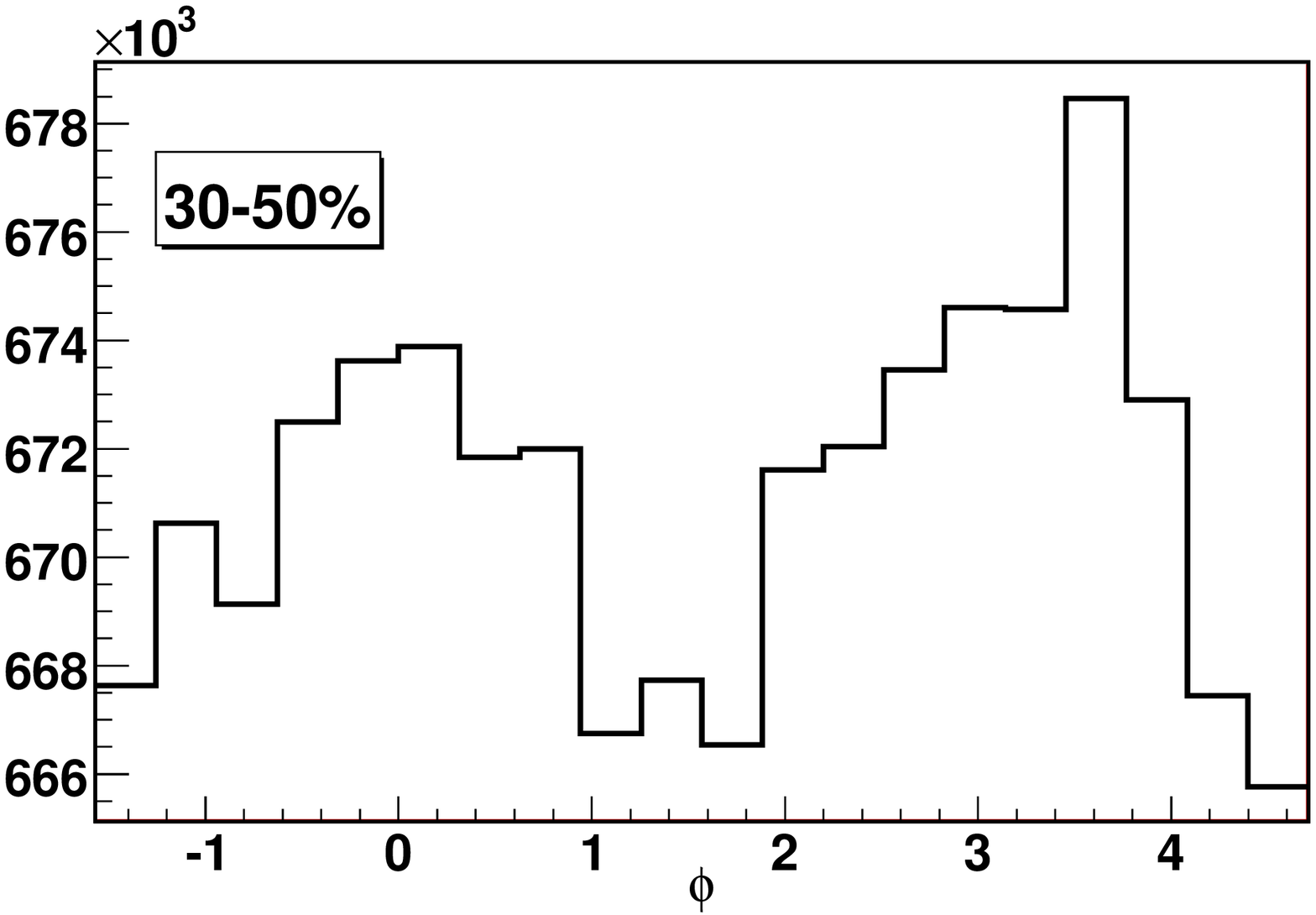}
\end{center}
\caption{%
The azimuthal distributions of produced pions  for the centrality classes 10--30\% (left)
and 30--50\% (right).}
\label{f:res}
\end{figure}

In Figure~\ref{f:res} the azimuthal distribution of pions produced in this model is shown.
%%%%%%%%%%%%%%%%%%%%%%%%%%%%%%%%%%%%%%%%%%%%%%%%%%%%%%%%%%%%%%%%%%%%%%
%%%%%%%%%%%%%%%%%%%%%%%%%%%%%%%%%%%%%%%%%%%%%%%%%%%%%%%%%%%%%%%%%%%%%
In total, 10,000 events were simulated. Histograms are shown for centrality 
classes 10--30\% and 30--50\%, where the azimuthal anisotropy of produced 
pion distribution is best seen. In both more and less central collisions the 
histogram is flatter or the statistics is worse. Though numerically the effect is small and gives $v_2$ 
of just slightly less than 1\% in this simulation, it is obvious from the histograms that the 
production asymmetry is correlated with the reaction plane. Note that the angle 
0 corresponds to the direction of the impact parameter. In other words, 
the angular distribution of pions produced in this model is given by the initial geometry
and \emph{cannot} be tagged as non-flow $v_2$ contribution.

%%%%%%%%%%%%%%%%%%%%%%%%%%%%%%%%%%%%%%%%%%%%%%%%%%%%%%%%%%%%%%%%%%%%%%

\section{Conclusions}

The numerical results presented here were obtained with a very simple toy model. 
Nevertheless, the message that they carry is that there could be a contribution 
to the elliptic flow from feeding in momentum from the  momentum loss 
of many hard partons which are expected at the LHC. It may be important to 
understand this contribution properly---together with all other effects determining 
the azimuthal asymmetry of the flow---if quantitative conclusions about the 
transport coefficients are to be drawn from the measurements. 

While we currently work on an improvement of the initial parametrisation of 
hard parton production, it goes far beyond the capability of the presented toy 
model to simulate proper response of the medium to momentum deposition from 
many hard partons. Interesting progress is being made in exploration of this problem, however. 
As argued here, it would be interesting to see a result of a proper hydrodynamic 
simulation of this situation.

%%%%%%%%%%%%%%%%%%%%%%%%%%%%%%%%%%%%%%%%%%%%%%%%%%%%%%%%%%%%%%%%%%

\acknowledgments

I thank the organisers for kind invitation to the workshop and the nice atmosphere created in Tokaj. 
I am grateful for discussions with P\'eter L\'evai, Fran\c cois Gelis, Evgeni Kolomeitsev, Urs Wiedemann, 
and Barbara Betz.

%%%%%%%%%%%%%%%%%%%%%%%%%%%%%%%%%%%%%%%%%%%%%%%%%%%%%%%%%%%%%%%%%%


\begin{thebibliography}{99}

\bibitem{Heinz:2001xi}
  U.~W.~Heinz and P.~F.~Kolb,
  %``Early thermalization at RHIC,''
  Nucl.\ Phys.\  A {\bf 702} (2002) 269
  [arXiv:hep-ph/0111075].

\bibitem{Pratt:2008sz}
  S.~Pratt and J.~Vredevoogd,
  %``Femtoscopy in Relativistic Heavy Ion Collisions and its Relation to Bulk
  %Properties of QCD Matter,''
  Phys.\ Rev.\  C {\bf 78} (2008) 054906
  [arXiv:0809.0516 [nucl-th]].

\bibitem{Broniowski:2008vp}
  W.~Broniowski, M.~Chojnacki, W.~Florkowski and A.~Kisiel,
  %``Uniform Description of Soft Observables in Heavy-Ion Collisions at
  %$\sqrt{s_{NN}} = 200$ GeV,''
  Phys.\ Rev.\ Lett.\  {\bf 101} (2008) 022301
  [arXiv:0801.4361 [nucl-th]].

\bibitem{Teaney:2003kp}
  D.~Teaney,
  %``Effect of shear viscosity on spectra, elliptic flow, and Hanbury
  %Brown-Twiss radii,''
  Phys.\ Rev.\  C {\bf 68} (2003) 034913
  [arXiv:nucl-th/0301099].

\bibitem{Stoecker:2004qu}
  H.~St\"ocker,
  %``Collective Flow signals the Quark Gluon Plasma,''
  Nucl.\ Phys.\  A {\bf 750} (2005) 121
  [arXiv:nucl-th/0406018].

\bibitem{Satarov:2005mv}
  L.~M.~Satarov, H.~St\"ocker and I.~N.~Mishustin,
  %``Mach shocks induced by partonic jets in expanding quark-gluon plasma,''
  Phys.\ Lett.\  B {\bf 627} (2005) 64
  [arXiv:hep-ph/0505245].

\bibitem{CasalderreySolana:2004qm}
  J.~Casalderrey-Solana, E.~V.~Shuryak and D.~Teaney,
  %``Conical flow induced by quenched QCD jets,''
  J.\ Phys.\ Conf.\ Ser.\  {\bf 27} (2005) 22
  [Nucl.\ Phys.\  A {\bf 774} (2006) 577]
  [arXiv:hep-ph/0411315].

\bibitem{Ruppert:2005uz}
  J.~Ruppert and B.~M\"uller,
  %``Waking the colored plasma,''
  Phys.\ Lett.\  B {\bf 618} (2005) 123
  [arXiv:hep-ph/0503158].

\bibitem{Betz:2008js}
  B.~Betz, M.~Gyulassy, D.~H.~Rischke, H.~St\"ocker and G.~Torrieri,
  %``Jet Propagation and Mach Cones in (3+1)d Ideal Hydrodynamics,''
  J.\ Phys.\ G {\bf 35} (2008) 104106
  [arXiv:0804.4408 [hep-ph]].

\bibitem{betz}
  B.~Betz, private communication, to be published.

\bibitem{Accardi:2004gp}
  A.~Accardi {\it et al.},
  %``Hard probes in heavy ion collisions at the LHC: Jet physics,''
  arXiv:hep-ph/0310274.

\end{thebibliography}
\end{document}